# The Coevolution of Banks and Corporate Securities Markets:

# The Financing of Belgium's Industrial Take-Off in the 1830s

Stefano Ugolini[1]

*Sciences Po Toulouse and LEREPS, University of Toulouse, Toulouse, France*

[1] Email: stefano.ugolini@ut-capitole.fr. I am grateful to Michelangelo Vasta and two anonymous referees for their great help in improving this article. I thank Olivier Accominotti, Vincent Bignon, Youssef Cassis, Abe de Jong, Marc Flandreau, Juan H. Flores, Timothy Guinnane, Lars Norden, Mary O'Sullivan, and Herman Van der Wee for discussing some of the issues raised here. I am also indebted to Frans Buelens for providing access to some data series from the SCOB database. An early version of this article was circulated under the title: "Universal Banking and the Development of Secondary Corporate Debt Markets: Lessons from 1830s Belgium". The usual disclaimers apply.

**Abstract:** *Recent developments in the literature on financial architecture suggest that banks and markets not only coexist, but also coevolve in ways that are non-neutral from the viewpoint of optimality. This article aims to analyse the concrete mechanisms of this coevolution by focusing on a very relevant case study: Belgium (the first Continental country to industrialize) at the time of the very first emergence of a modern financial system (the 1830s). The article shows that intermediaries played a crucial role in developing secondary securities markets (as banks acted as securitizers), but market conditions also had a strong feedback on banks' balance sheets and activities (as banks also acted as market-makers for the securities they had issued). The findings suggest that not only structural, but also cyclical factors can be important determinants of changes in financial architecture.*

**Section 1: Introduction**

Since the foundational contributions of Gurley and Shaw (1955), Gerschenkron (1962), and Goldsmith (1969), a massive body of research has been consecrated to the understanding of *financial architecture*, and the pros and cons of allegedly market- or bank-based financial systems have been thoroughly investigated. Starting from the late 1990s, the relevance of the traditional 'banks vs. markets' dichotomy has been questioned theoretically (Allen & Gale, 2000), empirically (Levine, 2005), as well as historically (Fohlin, 2012). A new consensus appears to have emerged on the fact that banks and markets should no longer be seen as substitutes, but rather as complements (Boot & Thakor, 2010). As a result, the question of the interaction between banks and markets has started to raise increasing interest. While the earlier literature had considered the evolution of banks and markets as a 'zero-sum game' in which the ones only grow at the expense of the others and vice-versa (Da Rin, 1997; Greenwood & Smith, 1997; Boot & Thakor, 1997; Boyd & Smith, 1998; Baliga & Polak, 2004; Chakraborty & Ray, 2007), more recent contributions have underlined that this need not be the case (Allen & Gale, 2000; Deidda & Fattouh, 2008; Mattana & Panetti, 2014). Their conclusion is that banks and market not only coexist, but also *coevolve* in ways that are non-neutral in terms of welfare. Banks perform three fundamental functions (screening, qualitative asset transformation, and credit enhancement) whose implementation allows for the development of financial markets; securitization puts in motion a virtuous cycle that is beneficial for the real economy, as it increases the efficiency of capital allocation (Song & Thakor, 2010).

Although economic historians have been paying increasing attention to the coexistence of banks and markets, the concrete mechanisms through which banks and markets coevolved in the past have not been specifically investigated to date. This article aims at filling this gap by studying the synchronous emergence of new intermediaries (universal banks) and new markets (secondary corporate stock markets) at the very time of the industrial take-off of a 'moderately backward' country (Belgium in the 1830s). The idea is to understand what factors drove the spectacular changes in financial architecture that allowed for the financing of growth in that national economic system, which was the first to follow Britain on the path to industrialization. As 1830s Belgium was the context in which universal banks first appeared at all, this article is also *de facto* a study of the very origins of universal banking.

The remainder is organized as follows. Section 2 reviews the historical literature on financial architecture. Section 3 surveys available historical evidence on Belgium and presents my original database. Section 4 focuses on the impact of banks' actions on market evolution. It underscores the role of Belgian banks in fostering the emergence of the Brussels stock market through their securitization of corporate assets. Section 5 focuses on the impact of market conditions on bank evolution. It shows that securitization did not mean a total dismissal of such assets by banks: banks actually had to continue intervening on the stock market well beyond the underwriting process, and securities originated by them tended to reappear on their balance sheets in times of crisis. Finally, Section 6 draws some general conclusions on the determinants of banks' and markets' coevolution.

**Section 2: Literature Review**

The historical literature on financial architecture has long been dominated by an "institutional" approach (Fohlin, 2016). On the one hand, Gerschenkron's (1962) highly influential contribution encouraged generations of historians to focus on the role of banks as drivers of industrialization in moderately-backward countries. On the other hand, Goldsmith's (1969) criticism of Gerschenkron led other historians to insist on the contribution of markets to the financing of economic growth. Both strands of the historical literature thus tended to take a (more or less strict) dicothomical view on the subject: financial systems were generally categorized as being *either* bank- *or* market-based. Yet, as argued by Fohlin (2016), the traditional "institutional" approach tends to obstruct a proper understanding of how financial systems actually worked, and should be replaced by a "functional" one. Focusing *ex post* on the provision of financial functions (rather than on the action of some specific organizations whose characteristics are defined *ex ante*) actually allows to by-pass the sterile controversies on the preeminence of banks or markets that have long dominated the historical literature.[2] In a series of contributions, Fohlin (2007, 2012, 2016) showed convincingly that financial architecture at the time of industrialization was far from being clearly bank- or market-based in any major country.

If recent contributions have substantially improved our understanding of how financial systems were structured at the time of industrialization, the very mechanisms through which such structures actually emerged and evolved over time still remain to be clarified. For instance, the best available synthesis of recent scholarship (Fohlin, 2012) adopts a cross-sectional rather than a truly chronological view on the subject. Even more detailed single-country studies (esp. Fohlin, 2007) are rather descriptive in nature and do not really ask *why* the provision of financial functions was actually organized the way it was. Yet, anecdotal evidence on the emergence of modern financial systems at the time of industrial take-off is actually available. In particular, three cases have attracted more scholarly attention: Germany, Austria, and Italy – the three countries that were identified by Gerschenkron (1962, p. 14) as paradigms of 'moderately backward' countries having adopted universal banking as a 'facilitator' of industrialization.

---

[2] For a discussion on the pros and cons of "institutional" and "functional" approaches, see Ugolini (2017).

Germany has naturally attracted much scholarly attention, but most of the focus has been on the mature stages of the financial system after the 1870s rather than on its infancy in the 1850s (see e.g. Gerschenkron, 1962; Fohlin, 2007). Available studies of the earlier phase of financial development (see e.g. Riesser, 1911; Tilly, 1966) have nonetheless shown that early German joint-stock banks' main business consisted of the floatation of 'innovative' companies on previously inexistent secondary corporate securities market.

Austria has also been mostly studied in the more mature stages of financial development since the 1890s rather than in the early times preceding the *Gründerkrach* of 1873 (see e.g. Gerschenkron, 1977; Good, 1991). Recent studies of this 'formative' crisis have however revealed the overwhelming involvement of early Austrian joint-stock banks into the underwriting of 'innovative' equities and into the related repo market, suggesting that banks were a driving force of the emergence of a stock market (Rieder, 2017).

In the case of Italy, the formative years of the modern financial system (the 1890s) have actually been covered by most of the literature (see e.g. Federico and Toniolo, 1991; Fohlin, 2012). These studies have consistently pointed to the same results found for 1850s Germany or 1870s Austria: in 1890s Italy, banks were largely concerned with the floatation of 'innovative' equities on the new stock exchange and provided liquidity to the latter through the repo market (Confalonieri, 1974-6; Brambilla, 2010).

Thus, available evidence on the very origins of financial architecture seems to suggest that early universal banks were heavily involved in the securitization of 'innovative' assets and played a crucial role in developing corporate securities markets. In what follows, this point will be thoroughly investigated through a case study on Belgium. Perhaps because of Gerschenkron's original oversight, this country has attracted comparatively little international attention. This lack of interest is unwarranted: Belgium was actually the first Continental country to experience, already in the 1830s, both an industrial take-off and the emergence of a modern financial system. More than that, 1830s Belgium was the place and time in which universal banking first emerged at all (Chlepner, 1943). The intermediaries that appeared in this context were very similar to the 'prototypical' universal bank: they extensively underwrote strictly industrial securities and collected deposits from the public (Ugolini, 2011). As Sylla (1991, p. 54) put it, 'Gerschenkron chose Germany as his example of a moderately backward country that employed dynamic banking to industrialize. He might better have chosen Belgium'.

## Section 3: Background and Data

At the beginning of the 19th century, the financial capital of the Belgian region was still located in Antwerp. The seat of Belgium's rich aristocracy, Brussels just hosted a number of tradesmen and a few bankers specialized in wealth management (Ugolini, 2011). Yet when in 1822 William I of the Netherlands decided to found a chartered bank of issue (acting as Treasury's agent) in the Southern part of his Kingdom, he chose to locate it in Brussels. Officially established to provide credit to infant industries (as suggested by its name: *Société Générale pour favoriser l'industrie nationale*, i.e. 'Financial Company for the Aid of the National Industry'), the bank was also intended for facilitating government borrowing in the South. The charter provided no restrictions on the financial operations the bank could implement, making it very different from the typical early-19th-century bank of issue (Demoulin, 1938, pp. 49-70).

Société Générale (hereafter SG) was supposed to be a purely private joint-stock company, but its stock capital was mostly provided by the King himself, who swapped a vast real-estate endowment (Crown forestlands) for equity. The rest was intended for sale to the public, but the rights issue failed, and William I found himself with a larger stake than planned (83%). The largest part of the remaining capital (a mere 9% of the total) was held by a group of Brussels notables (landowners, retailers, and private bankers), from whom the members of the board came (Brion & Moreau, 1998, pp. 23-24). In 1830, the directors of SG took advantage of the opportunity offered by the Belgian revolt: the King's stake was frozen, and Crown forestlands were seized as private assets of the company. *De facto*, the bank came to be owned by itself.

During the first decades of the 19<sup>th</sup> century, Belgium's industrial potential (tied to the presence of coal and iron ore deposits) had become increasingly clear, and a number of partnerships had been created with the aim of developing mechanized plants (Briavoinne, 1839; Mokyr, 1976). The new firms, however, were faced with the typical difficulties inherent with infant 'high-tech' sectors: high growth potential was actually counterbalanced by large uncertainty and by the volatility of revenues, thus discouraging lenders from providing credit (Lévy-Leboyer, 1964; Cameron, 1967). In the 1820s, credit rationing had been partially alleviated by direct state aid and government-sponsored lending by SG (Demoulin, 1938), but the 1830 revolt led to a 'sudden stop'. The crisis had two important consequences for the newborn country's financial system. First, municipal savings banks, which were heavily invested in Dutch public debt, went bankrupt, and SG was called to the rescue by absorbing them: as a result, the bank assumed large deposit-taking activities (Ugolini, 2011). Second, growth perspectives were jeopardized, and many infant industries slid into insolvency. Facing difficulties in recovering industrial credits, since late 1834 SG started to negotiate their conversion into equity. Industrial partnerships were thus transformed into joint-stock companies and then floated on the stock market, although the bank kept a stake in them. As argued by Briavoinne (1839, p. 230), the bank ventured into incorporation with the aim of *mobilizing* its own assets. The incorporation wave had started, and the first prototype of universal bank was born.

In 1834, a conflict burst between SG and the Belgian government, which was willing to seize the Dutch Crown's assets appropriated by the bank. In order to challenge SG's monopoly, in January 1835 a new joint-stock bank, called Banque de Belgique (hereafter BdB), was founded. The bank was intended to replace SG as the Treasury's agent, and therefore to become Belgium's main bank of issue – a step that the government did not eventually dare to take. BdB was designed along the model of SG in all details. This was due to the fact that, among its founders, there was a group of private bankers eyeing the possibility to restructure the underperforming debt of a number of firms to which they found themselves directly exposed (Chlepner, 1926, pp. 63-67). As a result, as SG began to float new companies on the Brussels stock exchange, BdB followed suit. In the space of a few months, their example was imitated by a number of privates unconnected with banks.

The Belgian incorporation boom came to a sudden stop in December 1838, when an exogenous shock (fear of a new war with the Netherlands) triggered bank runs. SG held out, while BdB suspended payments and was bailed out by the government. The effects of the crisis were long-lasting, as the Brussels stock market did not recover until the 1850s (Ugolini, 2011).

In what follows, these 'formative' events are analyzed on the basis of an original database including all available information on both markets and banks during the 1830s. Table 1 provides details on the series included in the database, as well as the sources from which they are drawn. The analysis of quantitative data performed in the following sections also builds on insights from a rich historical literature, including above all Chlepner's (1926, 1943) contributions, based on an impressive collection of qualitative sources.

[Table 1 about here]

## Section 4: From Banks to Markets

*4.1: Banks as Securitizers: Conceptual Framework*

The literature on the industrial organization of corporate finance generally maintains that banks are better positioned than markets for overcoming information asymmetries. This is due to the existence of economies of scope between lending and the provision of payment services, which allows bank to access relevant information on borrowers (Goodfriend, 1991). This explains why bank loans are often the only source of external funding available to firms at non-prohibitive costs.

But lending is not banks' sole corporate finance activity. Innovative start-ups, whose cash flows are highly uncertain, may need subsidized lending in their early stages, yet intertemporal cross-subsidization is only feasible if the lender-borrower relationship is durable (Petersen & Rajan, 1995). Lenders can solve this moral hazard problem by acquiring a stake in the borrowing firm and thus getting a say in future decision-making. Banks may hence behave as strategic investors in view of the synergies between venture capital investment and relationship lending (Hellmann et al., 2008). Debt restructuring may be another reason for banks to acquire equity in start-ups. Firms with a higher probability of default are more likely to approach banks, because of the latter's superior ability to minimize inefficient liquidations (Chemmanur and Fulghieri, 1994). Banks will be willing to accept a conversion of debt into equity when the market value of their claim on the firm is significantly lower than its fundamental value, when the firm's growth opportunities are high, and when indebtedness to other lenders is low (James, 1995). This means that debt-for-equity swaps are more likely to occur between banks and start-ups.

While banks may have good reasons to acquire equity in innovative start-ups, this does not mean they will necessarily keep it for long. Equity in unlisted firms is an idiosyncratic and illiquid asset. To reduce their exposure to liquidity risk, banks can perform qualitative asset transformation and convert it into standardized exchange-traded securities (Loutskina, 2011): the bigger the investment, the stronger the incentive to put it off balance sheet (Ghent & Valkanov, 2015). Securitizing past investments will allow banks to find new liquid resources for seizing new investing opportunities (Vargas-Martínez, 2010).

Empirical research has found a net certification effect by banks in the securitization process. Shares in companies underwritten by banks perform better than others: the effect is stronger when the bank keeps a stake in the firm, and is increasing in the size of the stake (Puri, 1999). The originator's continued involvement in the floated company is equivalent to an *IPO lockup* (Brav & Gompers, 2003): as it is supposed to ensure high-quality monitoring, it works as a credit enhancement for issued securities (Boot, 2000).

Three of the literature's findings are relevant for understanding the role of banks in the very early stages of industrialization. First, banks will have a predominant role in the funding of innovative start-ups both through lending (because of their privileged access to information on borrowers) and through capital venture investing (because of incentives to perform debt-for-equity swaps). Thus, banks can be expected to become the leading providers of capital to innovative firms in view of their superior *screening* technologies. Second, in order to continue profiting from their competitive advantage, banks will need to discharge part of the acquired equity by converting it into marketable securities. Thus, banks can be expected to promote the *qualitative transformation* of innovative corporate assets. Third, banks have a positive certification effect for newly-floated corporate securities that is direct proportional to their subsequent involvement into the floated company. Thus, banks can be expected to maintain corporate control on the floated companies, and this will work as a *credit enhancement* in the eyes of investors. These three hypotheses will now be tested in the context of Belgium's industrial take-off.

### *4.2: Banks as Securitizers: Evidence from 1830s Belgium*

At the time of Independence, the Belgian provinces possessed neither a developed capital market nor a sophisticated banking sector: the bourses of Antwerp and Brussels had thin trading activities in a handful of sovereign bonds, while the banking sector only consisted (if we exclude SG) of a bunch of small private banks unable to open sizable credit lines to industrial firms (Ugolini, 2011). In the 1820s, the earliest attempts at industrialization had thus been financed either through internal capital accumulation or through direct state aid (Mokyr, 1976). Tellingly, in the more than 250 pages dedicated by his treatise on Belgian industrialization to the 'institutions fostering economic growth', Natalis Briavoinne referred to banks only as providers of means of payments or as underwriters, but *not* as lenders (Briavoinne, 1839). The only exception was SG, but its intervention was seen as yet another form of state aid. In the 1820s, SG had been the one significant source of credit available to industrial concerns: as the Chamber of Commerce of Mons wrote in 1830, the local branch of SG had been "the main, and we can now say it with certitude, in the last analysis the one and only foundation" to the development of the coal industry in the region (Demoulin, 1938, p. 100; Ugolini, 2016). When the 1830 crisis erupted, the bank thus found itself largely exposed to defaults in the corporate sector (Houtman-De Smedt, 1997). Table 2 shows that in 1830-1, SG was indeed abnormally engaged into short-term lending in the Mons region, in order to meet the pressing demand of the corporate sector (especially coalmines). As soon as it became clear that loans would be unrecoverable, the bank started to negotiate debt-for-equity swaps with defaulting firms. The Mons region would thus become the heartland of SG's industrial empire (Brion & Moreau, 1998). Yet debt restructuring was not the main way in which banks took stakes in innovative firms. More typically, entrepreneurs approached their bank to ask funds for the modernization of their private partnership, and negotiations took place: then the firm was incorporated, a portion of the stock reserved for the entrepreneur, and the rest underwritten by the bank (Chlepner, 1943).

[Table 2 about here]

Following 1831, Belgium experienced a boom of incorporations. Under the aegis of the Napoleonic *Code de Commerce* (still in force in the country), three corporate forms were available at the time: 1) *société en nom collectif* (ordinary partnership with unlimited liability of all partners), 2) *société en commandite* (special partnership with unlimited liability of active partners and limited liability of silent partners), and 3) *société anonyme* (joint-stock company with limited liability of all shareholders). Because it provided an almost unchecked power to active partners and thus opened scope for abuses, *société en commandite* soon became a very unpopular corporate form. *Société anonyme* was considered as granting better protection to shareholders, but its creation had to be approved by decree, making it a hardly accessible corporate form for most industries (Freedeman 1965). The text of the new Constitution of 1831, however, opened a juridical loophole with non-negligible consequences for the corporate sector. The establishment of 'unlimited freedom of association' as a constitutional right was actually interpreted by many as an abolition of the requirement of government approval to the incorporation of *sociétés anonymes*. The result was an unchecked boom of creations of joint-stock companies (Chlepner, 1926). Figure 1 shows that the peak of the phenomenon took place between 1835 and 1838. As illustrated by table 3, almost two thirds of the total amount of capital raised went to three specific sectors: financials, coalmines, and ironworks. The role played by banks is highlighted by table 4: SG and BdB contributed to the incorporation of 37% of the companies created, and these very companies absorbed more than 60% of the total amount of capital raised. This means that banks acted as venture capital investors for more capital-intensive firms, whose growth prospects and cash flow volatility were highest at the time. On the whole, available evidence appears to be consistent with the first hypothesis: at the time of Belgium's industrial take-off, the provision of capital to innovative firms was actually dominated by banks.

[Figure 1 and Tables 3 and 4 about here]

Given the dearth of data concerning the Belgian capital market during the 1830s, the only available proxy for the width of this market consists of the number of securities exchanged at the bourse.[3] Table 5 describes the population of securities traded in Brussels by asset class. It shows that a Belgian stock market only emerged in 1835 (when universal banks entered the underwriting business), and continued to be dominated by banks throughout the period (always more than 80% of the total listed equities concerned bank-affiliated firms). The fact that a large number of incorporated firms did not have their stock traded at the Brussels bourse, however, is not imputable to the existence of entry barriers to outsiders (Chlepner, 1926). This suggests that the shares of non-affiliated companies, which tended to be less capital-intensive (see table 5), only circulated in restricted local circles (where information asymmetries could be lowered by personal connections: Lamoreaux, 1996), and lacked access to the official bourse. Thus, universal banks acted as gatekeepers of the general capital market. The non-existence of a market for corporate bonds corroborates this conclusion, as banks never underwrote such securities. This appears to confirm the second hypothesis – viz., that banks dominated the process of qualitative transformation of corporate assets into exchange-traded securities.

---

[3] Ugolini (2011) discusses the choice of this proxy and the reasons for preferring the lists provided by the press (which covered the securities actually traded on the market) to the lists provided by official bulletins (which covered all securities listed regardless of their actual liquidity). He also takes into account the other relevant Belgian capital market (i.e. the Antwerp bourse), to conclude that the flotation of industrial equity was a purely Brussels-based phenomenon.

[Table 5 about here]

According to the *Code de Commerce*, the names of the directors of new joint-stock companies had to be explicitly mentioned in the charter: the board typically included the promoters of the enterprise and was ratified by the shareholders' general assembly only well after the floatation (Freedeman, 1965). This arrangement allowed venture capital investors to keep their grip on the firm well beyond the incorporation process. Using information from the charters of floated companies, Table 6 shows that the directors of SG and BdB were systematically advertised as 'leading administrators' of the firms incorporated by the two banks. The governor of BdB, Charles de Brouckère, was indicated as 'leading administrator' of 13 companies underwritten by his bank, with a cumulative capital of more than 55 million francs; while the governor of SG, Ferdinand de Meeus, was either directly or indirectly (as a 'leading administrator' of the two subsidiaries Société de Commerce and Société Nationale) chairing no less than 24 companies underwritten by his bank, with a cumulative capital of more than 92 million francs. In the end, the boards of all bank-affiliated companies were constantly occupied by the directors of the two underwriters (Laureyssens, 1975; Kurgan-Van Hentenryk, 1996). Exacerbated by extensive crossholding of affiliates through other affiliates, the phenomenon of board interlocks proved persistent feature in Belgium (Van Overfelt et al., 2009). The fact that banks' corporate control on affiliated firms remained strongest is consistent with the idea of a lockup effect on underwritten companies. Banks' persistent involvement into the management of floated companies acted as a sort of moral guarantee in the eyes of investors, as it suggested that underwritten firms would be closely monitored and granted constant liquidity assistance by banks. On the whole, this seems to confirm the third hypothesis – viz., that banks boosted the attractiveness of floated securities by providing them with some form of credit enhancement.

[Table 6 about here]

**Section 5: From Markets to Banks**

*5.1: Banks as Market Makers: Conceptual Framework*

According to the finance literature, markets provide two crucial functions: *price discovery* and *liquidity*. The two are inextricably linked, although they do not actually coincide. On the one hand, price discovery is the mechanism allowing for the incorporation of all available information into asset prices: its efficiency is hindered by the presence of information asymmetries, as uninformed traders bear a non-diversifiable risk of losses when they trade against informed ones. On the other hand, liquidity is the mechanism allowing for the matching of buyers and sellers of assets: its efficiency is hindered by the presence of transaction costs (O'Hara, 2003).

Price discovery is a particularly delicate process at the time of the first introduction of an asset into the market. In theory, even if the amount of information available to the whole of market participants is superior to the one available to the underwriter alone, the latter's pricing errors should be randomly distributed – meaning that the difference between the issue price and the market price of the floated asset should be zero on average. In real-world markets, however, *underpricing* is almost constantly observed – meaning that the market price almost always exceeds the issue price. *Price run-ups* have been generally interpreted as unbiased indicators of information asymmetries – i.e., as the 'lemons premium' that issuers are obliged to pay in order for uninformed investors to be attracted (Rock, 1986). However, some contributions have pointed to the existence of widespread aftermarket intervention by underwriters – be it direct (such as price support) or indirect (such as sentiment creation) – which exacerbates price run-ups (Welch, 1992; Ellis et al., 2000; Derrien, 2005; Ritter, 2011).

Two (not mutually exclusive) explanations for aftermarket intervention have been advanced. The first (more pessimistic) one is that securitization is, by far, the main source of profit for venture capitalists (Lerner, 1994). If the same intermediary is both venture capitalist and underwriter of the floated firm, conflicts of interests will naturally arise in view of the obvious incentive to maximize price run-ups, and this may be conducive to cronyism (Drucker & Puri, 2007; Ritter, 2011). The second (more optimistic) explanation is linked to the sunk-cost nature of *reputation*. Success in the underwriting business critically depends on the underwriter's reputation as a certifier (Ljungqvist, 2007). In case the case of an IPO lockup, the underwriters' need to enhance the performance of the floated securities will not only apply in the short run, but also in the long run. In order to prevent bad signals from being conveyed to the market, underwriting banks will be ready to provide subsidized lending to an underperforming affiliated firm (Kanodia et al., 1989). Therefore, in order to maintain its reputation as an underwriter, a bank may find itself unable to eliminate *credit risk* through securitization – meaning that the qualitative asset transformation process is incomplete. Credit risk inherent to the originated securities might actually return to the bank under the form of informal off-balance-sheet commitments – as it spectacularly did to issuers of asset-backed securities in 2007 (Acharya et al., 2013).

The second fundamental function of markets consists of enhancing liquidity. Exchange-traded securities naturally tend to be more liquid than other assets in view of their standardized nature and of their access to a larger pool of potential buyers. Liquidity is one of the fundamental determinants of the attractiveness of securities (Amihud at al., 2005). However, liquidity can quickly evaporate in securities markets in the event of crises. As a result, the attractiveness of a financial asset is strongly increased by the presence of a *lender of last resort* (LLR) operating on the market for that specific asset. Through its commitment to lend on the asset at any moment at some price, the LLR provides a ceiling to transaction costs; in other words, the presence of a LLR acts as a guarantee against future liquidity shocks. Thus, assets covered by such a guarantee have a clear competitive advantage with respect to assets exposed to illiquidity risks, as they can be pledged to obtain credit at any moment. Central banks started enacting LLR policies in the mid-19$^{th}$ century, but for many decades they limited their action to a rather narrow range of securities – mainly high-quality bills of exchange and a limited number of sovereign and quasi-sovereign bonds (Bignon et al., 2012). This forced underwriters of corporate securities to step in: 19$^{th}$-century banks saw as a positive duty the commitment to lend on the securities they had themselves issued (Flandreau & Sicsic, 2001).

From what precedes, two insights can be drawn that are relevant for understanding corporate securities markets in the very early stages of industrialization. First, there are good reasons for underwriters to influence the price discovery mechanism – because profits can be made from price run-ups in the short run, and/or because reputation needs to be preserved in the long run. Thus, underwriting banks can be expected to actively implement *market interventions* in order to sustain the price and dividend performance of the underwritten companies, both in the short and in the long run. Second, there are good reasons for underwriters to sustain the liquidity of underwritten securities – because liquidity is an essential determinant of asset prices. Thus, underwriting banks can be expected to provide market participants with *lending of last resort* facilities on the underwritten securities. On the whole, this means that underwriters can be expected to behave as *market-makers* for the securities they issue. These hypotheses will now be tested in the context of Belgium's industrial take-off.

*4.2: Belgian Banks as Market Makers: Evidence*

The most popular measure of underpricing is the *IPO discount* – i.e., the difference between the issue price and the market price of the stock at the end of the first day of floatation. As data on IPO discounts are only available for four stocks (see Table 1), I compute an alternative measure of underpricing, which I call the *mid-term price run-up*: this is defined as the difference between the issue price and the mid-term equilibrium price of the stock (computed as the average price over the first six months in which the stock appears on bulletins). Table 7 compares mid-term price run-ups with actual IPO discounts for the only four stocks for which the latter are available. As expected, the two measures of underpricing differ, but they are of the same order of magnitude – except for one very special case.[4] The computed mid-term price run-ups are shown in Figure 2. Three general patterns can be observed: on average, SG-affiliated stocks displayed the higher run-ups; BdB-affiliated stocks displayed lower run-ups; and non-affiliated stocks generally saw their price fall. For all classes, run-ups decline in the years following 1835.[5]

---

[4] The price of Asphaltes Seyssel stock experienced a spectacular rise during some weeks, rapidly lost momentum, and then disappeared completely from listings: this gives a very high average price that cannot be considered as an equilibrium price.
[5] The conclusion does not seem to be valid for non-affiliated companies in 1838, but the result is entirely driven by the Asphaltes Seyssel stock (see footnote 4).

[Table 7 and Figure 2 about here]

Were run-ups due to *ex-ante* underpricing (as in Rock, 1986) or to *ex-post* intervention (as in Ritter, 2011)? Direct historical evidence does not allow establishing if and to what extent underwriters intervened in the aftermarket in order to support the price of the securities they issued. One indirect way to answer this question consists of determining whether stocks were overpriced in the mid run by comparing prices and actual returns. Data on dividends are only available for ten stocks (see table 1),[6] which are highly likely to suffer from the survivor bias – meaning that they supposedly cover only the high-quality end of the market. Past returns and prices for all assets on which information is available are displayed on Figure 3. Market prices appear to reflect the returns of the assets: the stocks whose price was higher actually paid higher dividends, and vice-versa; interestingly, the yield of corporate assets was more or less in line with the yield of Belgian government bonds. On the whole, no patently aberrant pricing of stocks can be found. This might be interpreted as pointing to the absence of a stock market bubble, and as suggesting that run-ups were due to *ex-ante* underpricing rather than to *ex-post* intervention by underwriters.

[Figure 3 about here]

Henceforth, the interpretation of underpricing as a 'lemons premium' necessary to attract outsiders to the securitization process does not seem to be disproved by *prima facie* evidence. However, the way the underwriting process was organized suggests that there was more to the story. As a matter of fact, initial subscribers were asked to pay only a small fraction of the nominal capital, and instalment payments were then spread across a long time span. Price run-ups thus provided a source of immediate gains for subscribers, who earned the difference between the issue price and the market price without having to put much capital on the table. In fact, the mechanism of allocation of new issues was very opaque. In the case of universal banks, bearers of affiliated securities were granted the right to subscribe new shares underwritten by the bank – which amounted to a sort of embedded call-option on future IPOs. But criteria for allocation were unclear, and insiders (typically, bank directors) managed to secure the lion's share in the business (Chlepner, 1926, pp. 87-91). Thus, directors had a clear incentive to multiply new IPOs: this may perhaps explain why the number of incorporations skyrocketed in the space of a few months. This sort of rent was extracted by subscribers at the expense of the underwriter: for newly-issued shares held on its own account, the bank was thus paid less than the aftermarket price. In modern parlance, the underwriter was '*spinning*' its own managers (Ritter, 2011). However, as directors had a clear incentive in maintaining the long-term viability of the mechanism and thus in *not* behaving as 'wildcat bankers', the rent was more akin to a tax than to a fully-fledged exploitation of the business by insiders.

---

[6] Available data cover eight SG-affiliated companies, two BdB-affiliated companies, and no non-affiliated company. SG-affiliated companies include: SG Holding Company (financials), HF Couillet (ironworks), Hornu et Wasmes (coalmining), Levant du Flénu (coalmining), Produits au Flénu (coalmining), Sars-Longchamps (coalmining), Sclessin (ironworks), Manufacture de Glaces (glassworks). BdB-affiliated companies include: BdB Holding Company (financials), Actions Réunies (financials).

The impression that bank directors did not behave as 'wildcats' is corroborated once we look at the long-run performance of the issued securities. At the time of industrialization, stocks were spurious securities: as firms' growth perspectives were highly uncertain, shares were often provided some enhancements, such as seniority over other liabilities or – more typically – a minimum dividend target (Baskin & Miranti, 1997). This is to be understood in the 19$^{th}$-century context, in which dividends played as a substitute for income statement transparency and thus had a paramount signaling role (Van Overfelt et al., 2010). In 1830s Belgium, equities were generally attached a yearly minimum dividend target equal to 5% of the issue price. As missing this target was considered somewhat like a default, underwriters that were keen on defending their reputation had an interest in avoiding the dividend underperformance of affiliated firms.

Table 8 reports available data on the dividend performance of bank-affiliated firms until 1848. Before the 1839 crisis, all companies largely beat the dividend target, thus conveying bullish signals to the stock market. During the economic depression of the 1840s, dividends declined for all firms – but according to different paths. On the one hand, all BdB-affiliated companies in the sample repeatedly missed the targets; on the other hand, only two out of the eight SG-affiliated companies in the sample did actually underperform. Such different paths might have been tied to the different strategies implemented by the two banks in the aftermath of the crisis. Following its bailout, in 1841 BdB was recapitalized by a group of new investors, who vowed to transform BdB into a pure commercial bank by liquidating its investment banking activities (Chlepner, 1926). As a result, the bank lost all incentives for defending its reputation as an originator of corporate securities, and consequently stopped performing subsidized lending to affiliated companies. Conversely, SG did not consider the opportunity of changing its business model, and therefore kept a strong interest in maintaining its reputational capital. The two SG-affiliated companies which repeatedly missed their targets in the 1840s (i.e. the Sclessin and Couillet ironworks) were among those more widely held by the bank herself (50.6% and 78.0% respectively), so that few investors were hit by the dividend cuts. Moreover, in order for the other SG-affiliated firms to avoid underperformance, the bank engaged into extensive subsidized lending and sponsored efficiency-enhancing company restructurings (Brion & Moreau, 1998). As long as they were willing not to quit the business, universal banks were thus ready to 'make sacrifices' in order for the present worth of their reputation not to drop – and this, even in a period of limited stock market activity as the 1840s (Ugolini, 2011). To sum up, while historical sources do not allow to find direct evidence of intervention by Belgian universal banks to sustain the price and dividend performance of floated companies, many pieces of indirect evidence strongly suggest that (consistent with the first hypothesis) active intervention did actually occur in the short as in the long run – at least, as long as banks attached a positive value to their reputation as securitizers.

[Table 8 about here]

I now turn to the question of liquidity provision. Different measures of liquidity have been proposed (Goyenko et al., 2009). In the simplest form, liquidity is estimated by looking at bid-ask spreads (a proxy for transaction costs): the higher the transaction costs, the lower the liquidity of the asset. In the case of emerging markets, however, even reconstructing bid-ask spreads is often an impossible task (Bekaert et al., 2007). This is also the case for 1830s Belgium. In order to overcome this problem, a viable alternative consists of focusing on *non-zero returns to equity* (Lesmond et al., 1999). The idea is straightforward: lack of price movements over time is interpreted as evidence of illiquidity. Figure 4 shows the number of securities listed on the Brussels stock exchange displaying non-zero week-on-week returns. It is possible to observe increasing liquidity up to the first months of 1838, and then a collapse in the summer of the same year. One element emerges from the picture: securities issued by universal banks were relatively liquid, while non-affiliated ones were utterly illiquid.

[Figure 4 about here]

Was this due to the intervention of universal banks themselves? For both banks, available information on securities accepted as collateral covers one single date only – viz. the day they were obliged to disclose their books to inspectors in the event of their bailout: December 13, 1838 for BdB, and March 1, 1848 for SG (tables 9.1 and 9.2).[7] It shows that universal banks almost exclusively lent on the securities they had underwritten themselves, while they only exceptionally took other securities (sovereign bonds included) as collateral. As confirmed by qualitative sources (Malou, 1863, p. 45), banks used to grant their affiliated securities eligibility for loans – i.e. they took the engagement of acting as LLR for them. Figures 5.1-2 give the total sums lent on securities by SG and BdB during the boom-and-bust cycle. They clearly show that as time passed, banks became more and more involved in meeting the demand for such loans. In the case of SG (for which yearly averages are available), lending on securities peaked in 1839; while in the case of BdB (for which only end-of-year figures exist), lending on securities was declining at the end of 1838 – when the bank had already fallen victim to the run. As the speculative wave was losing momentum, more and more shareholders were bringing their securities to the banks in order to obtain cash. This means that borrowing conditions on the market were worse than those offered by banks. As a result, from being the lenders of last resort, banks ended up being the *market-makers of last resort* for affiliated securities, as the market simply ceased to exist outside the banks. This explains why, on the one hand, the observed liquidity of BdB-affiliated securities evaporated once the run prevented BdB from continuing to perform its lending policies; and why, on the other hand, the observed liquidity of SG-affiliated securities was not completely impaired by the crisis, as SG continued to lend up to the point of 'absorbing' the whole market.[8] All this confirms the second hypothesis – viz., that banks offered lending facilities to market participants in order to sustain the performance of affiliated securities. On the whole, historical evidence confirms that banks played a primary role in the functioning of emerging markets by acting as market-makers for the securities they had themselves originated.

---

[7] Although data for SG concern a much later date than the events analyzed here, we know that most loans had been contracted back in 1839 and systematically rolled-over thereafter (Annales Parlementaires, 1848). We can thus expect these figures to be representative of SG's policy during the 1830s.

[8] The effect of this 'absorption' of the market by the bank was the complete immobilization of the assets side of its balance sheet: although SG managed to avoid suspending payments in 1839, its structure became extremely fragile and could not survive the following liquidity crisis in 1848 (Annales Parlementaires, 1848).

[Tables 9.1-2 and Figures 5.1-2 about here]

**Section 6: Conclusions**

Building on a substantial amount of primary and secondary qualitative and quantitative evidence, this article has provided detailed microeconomic evidence on the interaction between banks and markets in the early stages of the industrialization of a 'moderately backward' country. On the one hand, it has shown that Belgian banks played a substantial role in the creation of a stock market in Brussels in the 1830s. Before 1835, firms were rationed by the capital market and could only obtain funding from intermediaries (or the state). Banks acted as venture capitalists and arranged the conditions for the transformation of high-tech 'start-ups' into exchange-listed public corporations. However, banks did not discharge all their stakes in floated companies and continued to monitor and control them. As a result, the stock market emerged not as a competitor to banks, but – quite to the contrary – as an instrument for banks to facilitate and expand their corporate finance business. On the other hand, the article has shown that Belgian banks played a considerable role not only in creating, but also in managing the Brussels stock market. First, banks created the conditions for price run-ups to happen – not only as a strategy for attracting outsiders to the stock market, but also as a way for securing profits to insiders. Second, banks continued to provide subsidization to affiliated firms after their floatation, so that the assets that they had taken out of their balance sheet through securitization actually reappeared on their balance sheet during the crisis. Third, banks maintained the liquidity of issued securities by acting as lenders of last resort on them, and continued to provide liquidity to their holders during and after the stock market crisis of 1839 – thus ending up 're-absorbing' *de facto* most of the securitized assets. After having partially 'externalized' their corporate finance functions to the market in the 1830s, Belgian banks were therefore forced to 're-internalize' most of them during the 1839 crisis. This was due to the fact that despite initial success, in the long term banks failed to attract enough outsiders to the new trading floor. However, the fact that the securitization wave of the 1830s was brutally stopped and put to a stand-by in the 1840s does not mean that the methods followed by banks were necessarily fundamentally unhealthy. As a matter of fact, the very same methods were applied anew by Belgian universal banks since the 1850s, and then met a considerable success in expanding the corporate securities market and thus establishing Brussels as one of the most important financial centres in Europe (Van Overfelt et al., 2009; Ugolini, 2011).

These findings suggest that, in stark contrast to the traditional idea that markets are competitors to banks in the corporate finance sector, banks fostered the emergence of markets in order to facilitate and expand their own business. Markets and banks not only coexisted, but also coevolved: corporate assets were pushed out of intermediaries' balance sheets through securitization, but they also jumped back to their balance sheets by means of their aftermarket interventions. All this is consistent with the conclusions of the theoretical literature on the coevolution of banks and markets (esp. Song & Thakor, 2010), but also qualifies them under one important respect. Following the traditional approach, the literature continues to consider changes in financial architecture as the result of modifications in *structural* (long-term) conditions. Although in an increasingly cautious way, also economic historians have continued to focus more on the structural economic, political, or legal factors impacting institutional design rather than on the contingent factors impacting the industrial organization of finance (Fohlin, 2012, 2016). In this case study, intermediaries were found to alternatively 'externalize' and 'internalize' their corporate finance activities according to modifications not in structural, but in *cyclical* (short-term) conditions. Anecdotal evidence on the 'formative' phases of other national financial systems (1850s Germany, 1870s Austria, 1890s Italy) appears to be consistent with these findings on 1830s Belgium. Therefore, one might speculate that, for all its importance, the role of path dependence in shaping financial systems might perhaps have been overemphasized by economists and historians alike. Only a more systematic comparative analysis of the early emergence of financial architecture across countries will be able to tell us to what extent this is may actually have been the case.

**Tables**

| *Name of Series* | *Number of Series* | *Period of Coverage* | *Source* |
|---|---|---|---|
| Market price of selected securities traded on the Brussels bourse (end-of-week) | 48 industrial stocks; 2 government bonds | 1835-1839 | *Moniteur Belge*; *Journal du Commerce d'Anvers* |
| Number of securities traded on the Brussels bourse (end-of-year) | 6 classes of securities | 1834-1839 | *L'Indépendance Belge* |
| Sector of activity of joint-stock companies founded in Belgium and amount of capital raised | 151 companies | 1831-1839 | Briavoinne (1839) |
| Names and affiliation of leading administrators in Belgian joint-stock companies | 149 companies | 1833-1838 | Briavoinne (1839); Laureyssens (1975) |
| IPO discounts on the Brussels bourse | 4 companies | 1835-1838 | Chlepner (1926); Brion and Moreau (1998) |
| Dividends paid by companies listed on the Brussels bourse | 10 companies | 1835-1848 | SCOB database (University of Antwerp) |
| Total amount of industrial securities held and total amount of loans on securities by Belgian banks (end-of-year) | 2 banks | 1835-1842 | Chlepner (1926) |
| SG: Breakdown of corporate loans by region | 1 bank | 1830-1839 | De Troyer (1974) |
| SG: Average amount lent on securities during the year | 1 bank | 1835-1842 | Malou (1863) |
| SG: Breakdown of loans on securities by type of collateral | 1 bank, 1 date | 1848 | Annales Parlementaires 1848, CdR n° 251 (28th April 1848) |
| BdB: Breakdown of loans on securities by type of collateral | 1 bank, 1 date | 1838 | Archives Générales du Royaume (Brussels), Fonds Min. Finances, 307/1/15, A |

**Table 1:** Summary of series and sources in author's database.

|      | *Mons* | *Brussels* | *Rest of Belgium* | *TOTAL* |
|------|--------|------------|-------------------|---------|
| 1830 | 63.3   | 27.3       | 13.7              | *104.3* |
| 1831 | 20.5   | 35.9       | 0.0               | *56.4*  |
| 1832 | 8.9    | 24.6       | 1.3               | *34.8*  |
| 1833 | 6.0    | 22.6       | 1.6               | *30.2*  |
| 1834 | 5.9    | 22.5       | 5.5               | *33.9*  |
| 1835 | 4.2    | 35.4       | 26.0              | *65.6*  |
| 1836 | 5.0    | 27.3       | 25.4              | *57.7*  |
| 1837 | 2.3    | 25.9       | 6.0               | *34.2*  |
| 1838 | 0.8    | 24.1       | 7.3               | *32.2*  |
| 1839 | 0.3    | 27.5       | 6.7               | *34.5*  |

**Table 2:** Geographic breakdown of total amounts of new short-term corporate lending (i.e. discount of trade bills) accorded by SG, 1830-9 (in million francs). Source: De Troyer (1974, p. 110).

| Banks and Investment Trusts | 25.8% |
|-----------------------------|-------|
| Coalmining                  | 19.8% |
| Steelworks                  | 17.3% |
| Insurance                   | 4.3%  |
| Shipping                    | 3.2%  |
| Real Estate Financials      | 2.9%  |
| Flax Mills                  | 2.5%  |
| Printing                    | 2.5%  |
| Glassworks                  | 2.5%  |
| Sugar Refineries            | 2.3%  |
| Machinery                   | 1.9%  |
| Other                       | 15.0% |
| *TOTAL*                     | *100.0%* |

**Table 3:** Breakdown per industrial sector of the total amount of capital raised, 1831-9. Source: Author, from Briavoinne (1839, pp. 224-6).

|  | Number of Companies Founded | Amount of Capital Raised | Average Capital per Company |
|---|---|---|---|
| SG-Affiliated Firms | 31 | 102.0 | 3.3 |
| BdB-Affiliated Firms | 25 | 74.0 | 3.0 |
| Non-Affiliated Firms | 95 | 113.0 | 1.2 |
| *TOTAL* | *151* | *289.0* | *1.9* |

**Table 4:** Breakdown per bank affiliation of the total number of incorporations, amount of capital raised, and average capital per company (in million francs), 1831-9. Source: Author, from Briavoinne (1839, pp. 223-4).

|  | Domestic Sovereign and Subsovereign Bonds | Foreign Sovereign and Subsovereign Bonds | Corporate Bonds | Stocks of SG-Affiliated Firms | Stocks of BdB-Affiliated Firms | Stocks of Non-Affiliated Firms | TOTAL |
|---|---|---|---|---|---|---|---|
| 1834 | 4 | 13 | 0 | 1 | 0 | 0 | *18* |
| 1835 | 3 | 12 | 0 | 4 | 2 | 0 | *21* |
| 1836 | 4 | 10 | 0 | 19 | 9 | 4 | *46* |
| 1837 | 4 | 10 | 0 | 21 | 14 | 9 | *58* |
| 1838 | 5 | 9 | 0 | 21 | 17 | 8 | *60* |
| 1839 | 5 | 9 | 0 | 21 | 17 | 8 | *60* |

**Table 5:** Breakdown per asset class of the total number of securities traded at the Brussels stock exchange (end of year), 1834-9. Source: Author, from *L'Indépendance Belge* (1834-9).

| Names of People Mentioned as 'Leading Administrator' of a Belgian Joint-Stock Company in Briavoinne (1839) | Number of Companies | Cumulative Capital of Companies (million francs) |
|---|---|---|
| Charles de Brouckère (BdB) | 13 | 55.3 |
| Ferdinand de Meeus (SG) | 12 | 70.3 |
| « Gérants de la Société de Commerce » (leading adm.: F. de Meeus, SG) | 7 | 17.2 |
| John Cockerill (BdB) | 4 | 5.9 |
| « Gérants de la Société Nationale » (leading adm.: F. de Meeus, SG) | 4 | 4.9 |
| 9 other SG-related names (average) | 1.9 | 4.6 |
| 10 other BdB-related names (average) | 1.6 | 3.9 |
| 88 other names unrelated to SG or BdB (average) | 1.1 | 1.0 |

**Table 6:** Belgium, 1833-1838: Names of people mentioned as 'leading administrator' of a joint-stock company in Briavoinne's treatise on Belgian industrialization, number of companies of which the person is a 'leading administrator', and cumulative capital of the companies of which the person is a 'leading administrator'. Source: Briavoinne (1839, pp. 562-6); Laureyssens (1975).

| Company | Affiliation | IPO Discount | | Mid-Term Performance | |
|---|---|---|---|---|---|
| | | Date | Price Run-Up | Period | Price Run-Up |
| Société de Commerce | SG | Mar. 21, 1835 | 24% | Jan.-Jul.1836 | 31.15% |
| Sars-Longchamps | SG | Nov. 4, 1835 | 14% | Jul.1836-Jan.1837 | 11.56% |
| Raffinerie Nationale | SG | Jun. 19, 1836 | 12% | Jan.-Jul. 1837 | 18.04% |
| Asphaltes Seyssel | None | Feb. 10, 1838 | 10% | Mar.-May 1838* | 39.92%* |

**Table 7:** Brussels bourse, 1835-1838: IPO discounts and mid-term price run-ups of four selected stocks. Source: Chlepner (1926, p. 92); Brion and Moreau (1998, p. 62); author's database.

|  | SG | | | BdB | | |
|---|---|---|---|---|---|---|
|  | *Excess Dividends with respect to the Target (% of the Stock's Issue Price): Average* | *Number of Companies in the Sample* | *Of Which Missing the Target* | *Excess Dividends with respect to the Target (% of the Stock's Issue Price): Average* | *Number of Companies in the Sample* | *Of Which Missing the Target* |
| 1835 | 0.95% | 1 | 0 | - | 0 | 0 |
| 1836 | 1.52% | 4 | 0 | 1.10% | 1 | 0 |
| 1837 | 1.69% | 7 | 0 | 2.80% | 1 | 0 |
| 1838 | 3.50% | 8 | 0 | 2.39% | 2 | 0 |
| 1839 | 2.18% | 8 | 0 | 0.00% | 2 | 0 |
| 1840 | 1.35% | 8 | 1 | 0.00% | 2 | 0 |
| 1841 | 0.90% | 8 | 1 | 0.00% | 2 | 0 |
| 1842 | -0.10% | 8 | 2 | -1.75% | 2 | 2 |
| 1843 | -0.20% | 8 | 2 | -1.43% | 2 | 2 |
| 1844 | -0.36% | 8 | 2 | -1.63% | 2 | 2 |
| 1845 | -0.04% | 8 | 2 | -2.08% | 2 | 2 |
| 1846 | 1.84% | 8 | 1 | -0.80% | 2 | 2 |
| 1847 | 1.26% | 8 | 2 | 0.95% | 2 | 0 |
| 1848 | 1.30% | 8 | 1 | -0.50% | 2 | 1 |

**Table 8**: Dividend performance of bank-affiliated companies, 1835-48. Source: Author, from SCOB database.

| Own stock | 16,95% |
|---|---|
| Main 4 affiliated 'financials' | 32,87% |
| Main 3 affiliated 'collieries and ironworks' | 11,27% |
| Main 3 affiliated 'other sectors' | 13,41% |
| Undetermined | 25,51% |

**Table 9.1:** SG, 1st March 1848: Breakdown of total sums lent on securities (by classes of securities). Source: author's computations on Annales Parlementaires (1848).

| Own stock | 2,12% |
|---|---|
| Listed affiliated 'financials' | 24,68% |
| Listed affiliated 'collieries and ironworks' | 34,22% |
| Listed affiliated 'other sectors' | 9,74% |
| Générale group stock | 1,54% |
| Listed non-affiliated stock | 9,36% |
| Unlisted stock | 2,14% |
| Sovereign bonds | 16,19% |

**Table 9.2:** BdB, 13th December 1838: Breakdown of total sums lent on securities (by classes of securities). Source: author's computations on AGR Brussels, Fonds Min. Finances, 307/1/15, A.

**Figures**

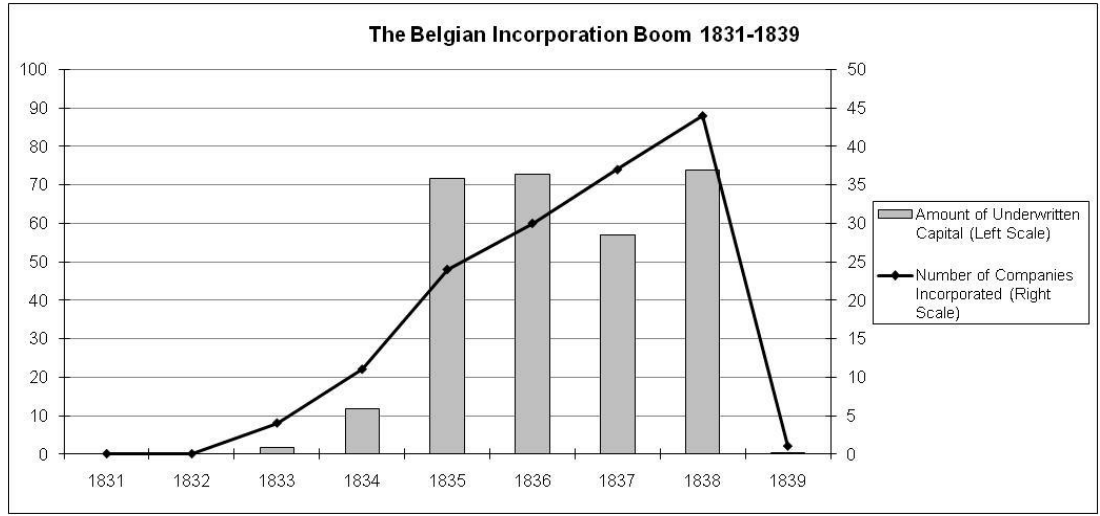

**Figure 1:** Total number of joint-stock companies founded (right scale) and total underwritten capital (in million francs, right scale), 1831-9. Source: Briavoinne (1839, pp. 223-224).

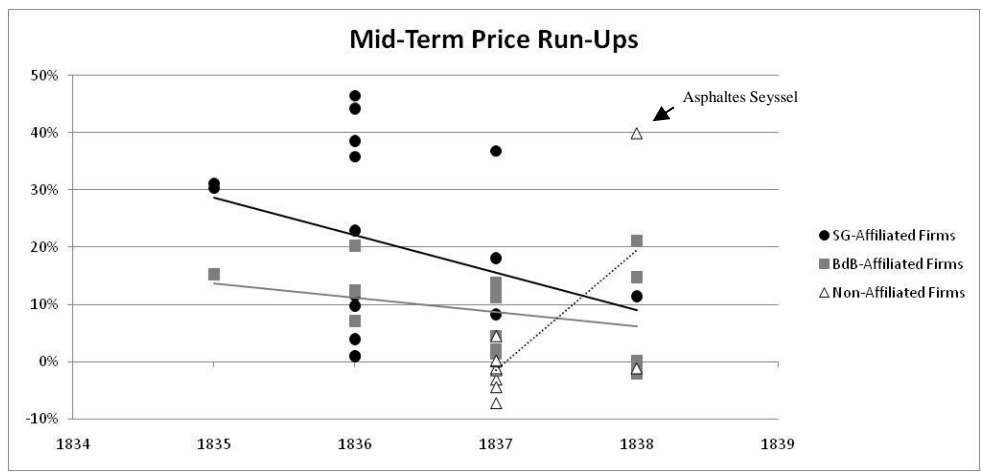

**Figure 2:** Mid-term price run-ups (defined as the difference between issue price and the average market price during the first six months of presence on the bulletins), 1834-9. Source: author's database.

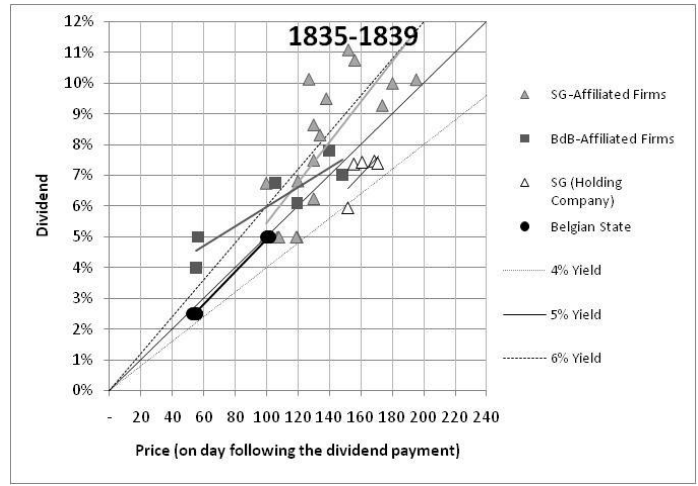

**Figure 3**: Total dividend yields of a number of securities by affiliation, 1835-9. Source: author's database; SCOB database.

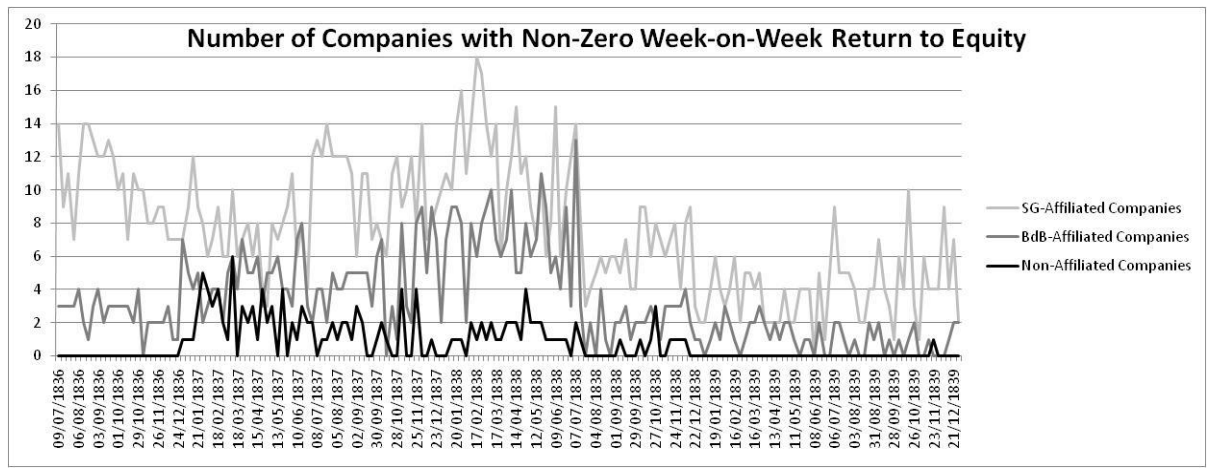

**Figure 4:** Number of companies with non-zero week-on-week return to equity (by affiliation), 1836-9. Source: author's database.

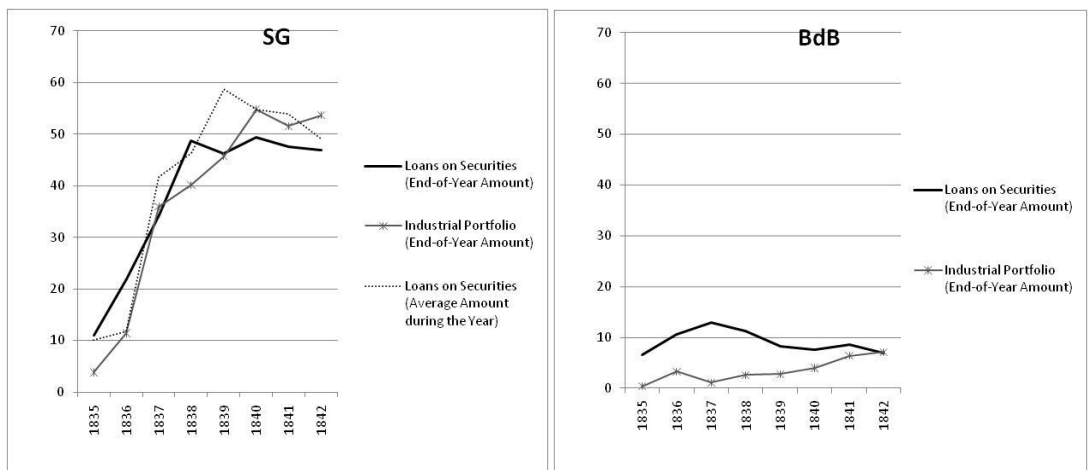

**Figures 5.1-2:** Loans on securities and industrial portfolio by Belgian banks (end-of-year amount, in million francs), 1835-1842. Source: Malou (1863, pp. I-IV); Chlepner (1926, pp. 78-79). SG, 1835-1842: Average amount lent on securities during the year (in million francs). Source: author's computations on Malou (1863, p. XX) and Annales Parlementaires (1848).

42